\documentclass[sigconf,edbt]{acmart-edbt2025}

\def\BibTeX{{\rm B\kern-.05em{\sc i\kern-.025em b}\kern-.08em
    T\kern-.1667em\lower.7ex\hbox{E}\kern-.125emX}}

\usepackage{booktabs} % For formal tables

\setcopyright{rightsretained}

\acmDOI{}

\acmISBN{978-3-98318-097-4}

\acmConference[EDBT 2025]{28th International Conference on Extending Database Technology (EDBT)}{25th March-28th March, 2025}{Barcelona, Spain}
\acmYear{2025}

\usepackage[utf8]{inputenc}
\usepackage[toc,page]{appendix}
\usepackage[export]{adjustbox}
\usepackage{algorithm} 
\usepackage[noend]{algpseudocode}
\usepackage{amsmath}
\usepackage{multirow}
\usepackage{tikz}
\usepackage{xcolor}

\DeclareMathOperator*{\argmax}{arg\,max}

\usepackage{todonotes}
\newcommand{\rev}[1]{\textcolor{black}{#1}}

\title{GraLMatch: Matching Groups of Entities with \underline{Gra}phs and \underline{L}anguage \underline{M}odels}

\author{Fernando De Meer Pardo}
\email{fernando.demeerpardo@zhaw.ch}
\affiliation{
    \institution{University of Zurich}
    \city{Zurich}
    \country{Switzerland}
}
\affiliation{
    \institution{Zurich University of Applied Sciences}
    \city{Winterthur}
    \country{Switzerland}
}

\author{Claude Lehmann}
\email{claude.lehmann@zhaw.ch}
\affiliation{
    \institution{Zurich University of Applied Sciences}
    \city{Winterthur}
    \country{Switzerland}
}

\author{Dennis Gehrig}
\email{dennis.gehrig@zhaw.ch}
\affiliation{
    \institution{Zurich University of Applied Sciences}
    \city{Winterthur}
    \country{Switzerland}
}

\author{Andrea Nagy}
\email{andrea.nagy@movedigital.ch}
\affiliation{
    \institution{Move Digital AG}
    \city{Zurich}
    \country{Switzerland}
}

\author{Stefano Nicoli}
\email{stefano.nicoli@movedigital.ch}
\affiliation{
    \institution{Move Digital AG}
    \city{Zurich}
    \country{Switzerland}
}

\author{Branka Hadji Misheva}
\email{branka.hadjimisheva@bfh.ch}
\affiliation{
    \institution{Bern University of Applied Sciences}
    \city{Bern}
    \country{Switzerland}
}

\author{Martin Braschler}
\email{martin.braschler@zhaw.ch}
\affiliation{
    \institution{Zurich University of Applied Sciences}
    \city{Winterthur}
    \country{Switzerland}
}

\author{Kurt Stockinger}
\email{kurt.stockinger@zhaw.ch}
\affiliation{
    \institution{Zurich University of Applied Sciences}
    \city{Winterthur}
    \country{Switzerland}
}
\renewcommand{\shortauthors}{De Meer Pardo et al.}

\begin{document}

\begin{abstract}
In this paper, we present an end-to-end multi-source Entity Matching problem, which we call \emph{entity group matching}, where the goal is to assign to the same group, records originating from multiple data sources but representing the same real-world entity. \rev{We focus on the effects of \emph{transitively matched records}, i.e. the records connected by paths in the graph $G = (V,E)$ whose nodes and edges represent the records and whether they are a match or not}. We present a real-world instance of this problem, where the challenge is to match records of companies and financial securities originating from different data providers. We also introduce \emph{two new multi-source benchmark datasets} that present similar matching challenges as real-world records. A distinctive characteristic of these records is that they are regularly updated following real-world events, but updates are not applied uniformly across data sources. This phenomenon makes the matching of certain groups of records only possible through the use of transitive information. 

In our experiments, we illustrate how considering \rev{\emph{transitively matched records}} is challenging since a limited amount of false positive \rev{pairwise} match predictions can throw off the group assignment of large quantities of records. Thus, we propose \emph{GraLMatch}, a \emph{method that can partially detect and remove false positive \rev{pairwise} predictions through graph-based properties}. Finally, we showcase how fine-tuning a Transformer-based model (DistilBERT) on a reduced number of labeled samples yields a better final entity group matching than training on more samples and/or incorporating fine-tuning optimizations, illustrating how precision becomes the deciding factor in the \rev{entity group} matching of large volumes of records.
\end{abstract}

\maketitle
\section{Introduction}\label{Section:Introduction}

\begin{figure*}
    \centering
    \includegraphics[width=0.8\linewidth, keepaspectratio]{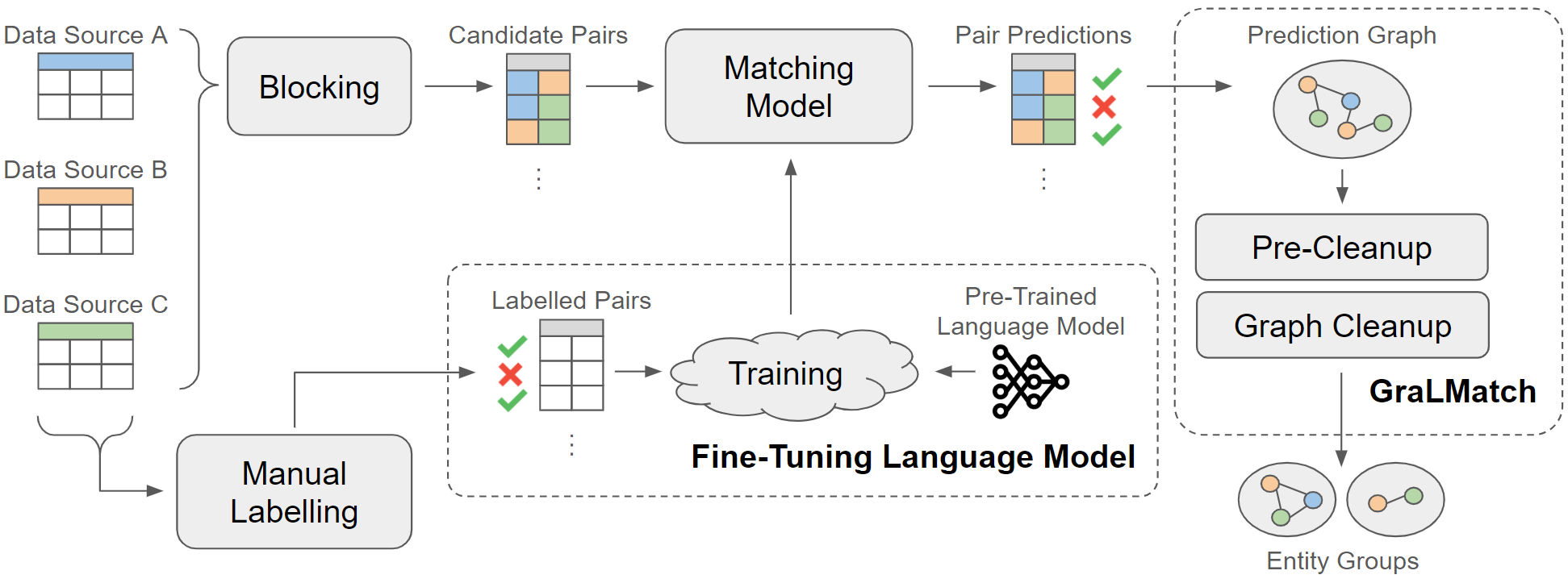}
    \caption{Illustration of the workflow of our entity group matching methodology.}
    \label{fig:entity group matching Workflow Diagram}
\end{figure*}

Due to remarkable technological advancements in recent decades, data has assumed a pivotal role within today's information society. As the volume of data experiences exponential growth \cite{reinsel2017data}, data integration, the process of aggregating and cleaning data originating from different sources, has become highly demanding. In fact, the most time-consuming aspect of the entire data science pipeline is often the cleaning and preparation of data \cite{deng2017data, holzer2022detecting}.

A vital part of data integration is \emph{Entity Matching} (EM). EM describes the problem of determining whether two or more records refer to the same real-life entity. If both records have the same structure i.e., share unique identifiers, this is a trivial task. However, in the instances in which the structure differs significantly, the EM tasks become increasingly complex. Earlier approaches to address this challenge rely on rule-based algorithms and heuristics \cite{Magellan2016, JedAI2020}. More recently, state-of-the-art systems rely on Machine Learning (ML) systems and are Transformer-based \cite{2017AttentionTransformer, Brunner2020}.

EM is a computationally intensive  process, as given the matching of $n$ records there are ${n\choose 2}$ possible pairwise matches to consider. Most state-of-the-art systems and benchmarks perform matching of $2$ data sources and thus focus on evaluating pairs of records as potential matches. In general, however, Entity Matching requires considering not only a set number of record pairs but also all \emph{transitively matched} records. 

We consider records $r_i$ and $r_j$ to be \emph{transitively matched} by a given \rev{pairwise} matching logic if there exists a path between them in the graph $G = (V,E)$ whose nodes and edges represent the records to be matched and the predicted \rev{pairwise} matches of a \rev{pairwise} matching logic respectively. All \emph{transitively matched} records are implied to be matches. The expected output of a matching is thus a list of \emph{groups of records} represented as \emph{complete graphs} whose nodes and edges represent records and matches (both predicted and transitive), respectively. 

\rev{Note that transitively matched records can appear in \emph{any matching setting}, no matter the number of data sources involved. Consider, for example, the matching of data sources S1 and S2 with records A \& B belonging to S1 and record C belonging to S2. If a pairwise matching logic predicts matches [A-C] and [B-C], then records A \& B will be transitively matched, i.e. the transitive match [A-B] is implied by the two previous pairwise matches. Current pairwise matching approaches do not take transitive matches into account and thus ignore the group assignments implied by their pairwise predictions.}

In this paper, we refer to the end-to-end multi-source EM problem as \textit{entity group matching}, and \rev{propose a novel matching methodology that combines state-of-the-art pairwise EM models and \emph{GraLMatch}, a Graph Cleanup technique that detects and removes false positive pairwise predictions to produce groups of records}. Starting from the raw source tables, we apply a blocking to select a subset of candidate record pairs that we predict as either matches or non-matches through some \rev{pairwise} matching model. At this stage we apply GraLMatch, a Graph Cleanup method that processes all the pairwise predictions, removes those likely to be false positives and produces groups of records. Figure \ref{fig:entity group matching Workflow Diagram} provides a visual representation of the entity group matching methodology we propose, which is not limited to language model-based \rev{pairwise} matching models, but also supports any matching method that produces pairwise matches.

We focus on a real-world instance of \emph{entity group matching}, where the challenge is to match records of companies and financial securities. The records we want to match are provided by financial data vendors and thus represent licensed information which cannot be readily shared. To overcome this, we generate two \textit{new multi-source synthetic benchmark datasets} that aim to mimic the characteristics of the original financial records. Starting from a publicly available set of records, we carry out a series of algorithmic modifications in order to create groups of companies and securities records that present similar matching difficulties as real-world examples.

A distinctive characteristic of the datasets that we target with our methodology compared to other commonly used benchmarks for EM is that their records are regularly updated reflecting real-world events\footnote{Such as rebrandings, mergers $\&$ acquisitions, new stock listings, bankruptcies, etc.} that may not be recorded equally across data sources. This \textit{data drift} phenomenon, along with instances of missing data, makes some matches between altered records impossible to identify based on their individual attributes alone. As stressed before, this phenomenon calls for methods capable of leveraging transitive information among records.

\begin{figure*}[h!t]
  \centering
  \includegraphics[width=0.8\textwidth, keepaspectratio]{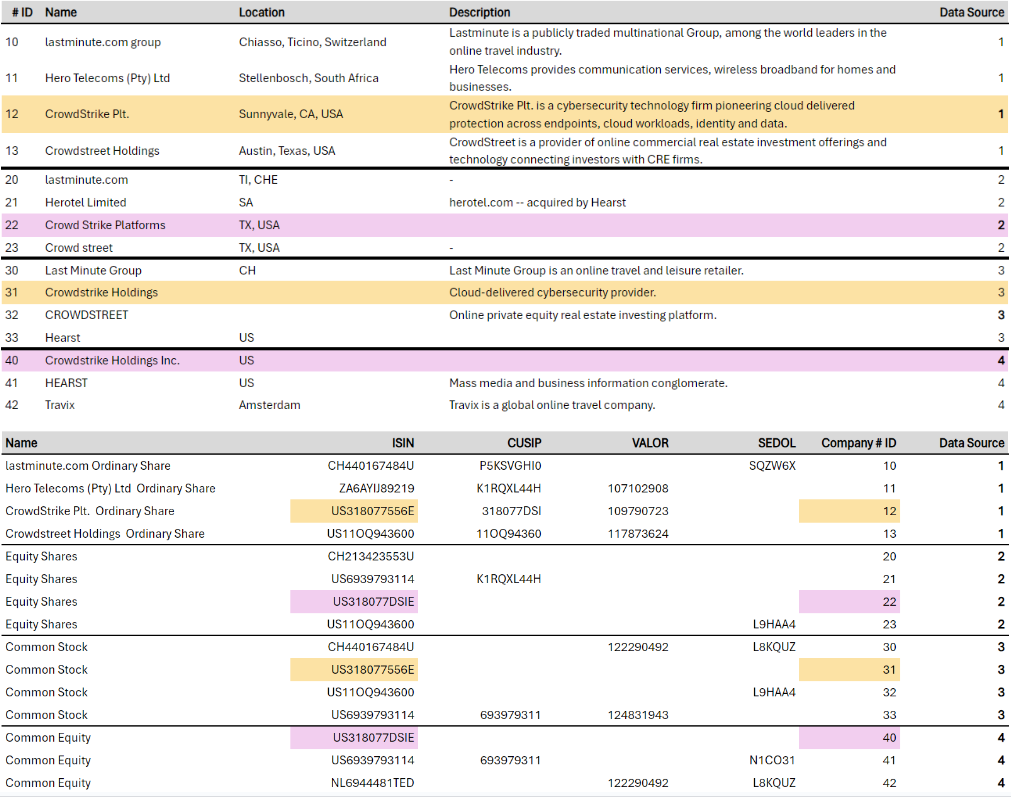}
  \caption{An example dataset of companies (top part) and securities records (bottom part) to match across multiple data sources. Records \#12, \#22, \#31 and \#40  correspond to the same entity, "Crowdstrike". Record \#12 can be matched to \#31 because they have securities with a matching ISIN, US31807756E highlighted in orange. Equivalently with \#22 and \#40 with US318077DSIE highlighted in violet. Matching the entire group however, requires recognizing all of the different naming variations as equivalent (Crowdstrike Plt./Crowd Strike Platforms/ Crowdstrike Holdings etc.). This task is not trivial, since many false positive predictions are likely to happen with, for example, records \#13, \#23, \#32 corresponding to the entity "Crowdstreet", due to the long shared character sequences across records.
  }
  \label{fig:matching_across_data_sources}
\end{figure*}

An illustration of the entity group matching problem is given in Figure \ref{fig:matching_across_data_sources}  where the companies and securities records come from four different data sources. These record groups illustrate some of the different matching challenges present both among real records and our synthetic benchmarks. Some record groups can be matched via the identifiers of their corresponding securities. However, matching identifiers do not guarantee a correct match due to \textit{data drift}. Other groups need to be matched by \textit{text alignment}, i.e. recognizing records as matches via their textual attributes. However, trying to find such groups unavoidably leads to false positive matches and thus to incorrect \emph{transitively matched records}. We deal with these false positives via \emph{GraLMatch.} 

\rev{Overall, the aim of this paper is to introduce the \emph{entity group matching} task and the concept of \emph{transitively matched} records. We evaluate the performance  of state-of-the-art pairwise EM algorithms on challenging datasets, illustrate the effect of transitively matched records and investigate whether our novel algorithm GraLMatch, can be combined with the pairwise algorithms to achieve good entity group matching results.} Consequently, the main \underline{contributions} of our paper are as follows:

\begin{enumerate}
    \item \rev{We introduce the entity group matching task and illustrate its comparatively higher difficulty w.r.t pairwise matching due to transitively matched records. In our experiments, we illustrate how the evaluations of pairwise EM algorithms carried out in previous works are lacking in terms of entity group matching. To the best of our knowledge, this is the first paper to introduce the problem of matching entity groups, and thus lays the foundations for novel research challenges.}
    \item \rev{We introduce two new challenging synthetic multi-source benchmark datasets for entity group matching, inspired by a real use case.}
    \item \rev{We present GraLMatch, a method that addresses the transitivity problem and uses graph-based techniques to detect and remove false positive pairwise predictions that lead to incorrect transitively matched records.}
\end{enumerate}

\rev{The remainder of the paper is organized as follows. In Section 2 we discuss the related work. Section 3 introduces the synthetic datasets we generate and the real-world use case that inspires them. Section 4 describes the entity group matching problem with a focus on \emph{transitively matched records} and our proposed methodology to deal with them. Section 5 describes the datasets, models and the scores of our experimental setup. Section 6 discusses the results of the experiments and Section 7 outlines our conclusions.}
% \begin{enumerate}
%     \item Previous work has mainly focused on matching \underline{pairs of entities}. We introduce \emph{GraLMatch - a novel matching methodology that combines Transformer-based language models and graph techniques to match \underline{groups of entities}}. To the best of our knowledge, this is the first paper to introduce the problem of matching entity groups, and thus lays the foundations for novel research challenges.
%     \item We provide  \emph{two challenging multi-source synthetic datasets to study entity group matching} based on real data from the financial industry. 
%     \item We showcase how \textit{entity group matching} is a significantly more challenging task than \emph{pairwise matching}, since very few false positive predictions can lead to a high number of \emph{false transitive matches}, which signify incorrect group assignments. Our experiments show how this issue can be mitigated via graph-based methods, such as \emph{GraLMatch}.
    
% \end{enumerate}

\section{Related Work}\label{Related Work}

Early approaches to EM tackled the problem through combinations of hand-crafted features and rule-based heuristics \cite{Magellan2016, JedAI2020} that struggle when faced with dirty/unstructured records. The advent of NLP techniques, more specifically Transformer models, has led to the current state-of-the-art EM systems which frame EM as a ML problem and treat records as instances of textual data.

Transformer Language Models are a class of neural network models based on the Transformer architecture \cite{2017AttentionTransformer} that have been shown to excel at numerous NLP tasks \cite{OverviewTransf2020}. These models are pre-trained on large text corpora in order to acquire natural language understanding via self-supervised tasks. In order to perform downstream tasks, one can perform Transfer Learning by adding task-specific final layers to the Transformer model initialized with the pre-trained weights and then fine-tune it on a task-specific training set until convergence, usually for a small number of epochs.
% such as Masked Language Modelling (MLM)\footnote{MLM is a prediction problem in which the model is tasked with predicting a subset of tokens of a given sentence that have been swapped with a special \textit{[MASK]} token.} and Next Sentence Prediction (NSP)\footnote{NSP is a binary classification problem in which given two sentences \textit{A} and \textit{B}, the model is tasked with classifying the pair as \texttt{IsNext} if \textit{B} follows \textit{A} in the original text or as \texttt{NotNext} otherwise.}.

BERT (Bidirectional Encoder Representations from Transformers) and its variants \cite{devlinBERT2018pretraining, sanh2019distilbert, RoBERTa2019} are a class of models consisting of the Encoder part of the Transformer architecture. They have been shown to achieve state-of-the-art \rev{pairwise matching results at numerous EM benchmark datasets} by following the fine-tuning process mentioned above along with different optimizations specific to the EM task such as Data Augmentation and Contrastive Learning \cite{Brunner2020, Li2023, DLforBlockinginEM2021, miao2021rotom, Peeters2022SupervisedCL, wang2022sudowoodo, Yao2022EntityRW} that aim to tackle the specific challenges presented by EM: (1) Lack of labeled training pairs, and (2) the difficulty of obtaining quality embeddings for nuanced similarity-based discriminative tasks, given the ambiguity present between many records. Additionally, optimizations explored for \textit{multi-source Entity Matching} include combining binary and multi-class classifications \cite{JointBert2021} and active learning using the graph implied by pairwise matches for query selection and data augmentation \cite{ALMSER2021}. 

\rev{All previous works, however, only evaluate models according to their performance at pairwise matching, ignoring transitive matches. Compared to the setting with only 2 data sources, transitive matches can be expected to have a greater effect in matching settings with multiple data sources\footnote{Simply because greater numbers of candidate matches are evaluated and thus false positive pairwise predictions are more likely.}. However, as we discussed in the previous section, transitively matched records can appear in \emph{any matching setting}. Previous models have only been evaluated according to their performance on pairwise matching in set data splits, which is a much easier task than entity group matching. In our experiments, we will show how a close-to-perfect pairwise matching performance is not sufficient to achieve a good matching if transitive matches are not considered.}

More recently, Transformer architectures have been scaled to billions of parameters, leading to a new class of models known as \emph{Large Language Models} (LLMs). These models are trained in the same self-supervised way as their smaller counterparts but are additionally further tuned via \emph{Reinforcement Learning with Human Feedback} (RLHF), which incorporates human input into the training process in order to align the model's behaviour with human preferences. LLMs have been shown to be capable of carrying out pairwise matching by framing it as a question-answering problem \cite{2022FoundModelsEM, 2023ChatGPTEM}. While promising, these models are considerably slower at inferencing and thus unsuitable for matching large datasets such as the ones we propose, as this requires millions of \rev{pairwise match evaluations} to be carried out.

Many benchmark datasets have been used over the years to compare the performance of newly proposed EM methodologies against existing ones. Older benchmarks mostly contain structured records of 2 data sources \cite{Kpcke2010EvaluationOE,Mudgal2018DeepLF, Draisbach2010DuDeTD}, while more recent ones often present records consisting only of textual descriptions originating from more data sources, which go into the thousands for web-scraped records \cite{Peeters2024WDCPA}. Additionally, multiple Data Generators that produce datasets of varying difficulty by polluting certain records (e.g. via modification or deletion of certain attribute values) have been also proposed  \cite{Wang2021MachampAG, Saveta2015LANCEPT, Christen2013FlexibleAE, Ferrara2011BenchmarkingMA, Hildebrandt2020LargeScaleDP, Primpeli2022ImpactOT, Ioannou2013}. We add to this line of research by proposing a multi-source matching problem that requires leveraging transitive information.

\section{entity group matching Problem: A Real-World Use Case}\label{Section: Syn Dataset Generation}

In this section, we introduce a use case of entity group matching based on the integration of financial records originating from various real-world data sources. This section serves as the basis for the generation of \emph{two new entity group matching benchmark datasets}. First, we describe the origin and nature of the \emph{original data}, which, due to privacy \rev{regulations}, cannot be shared with the research community. Next, we describe the generation of the \emph{synthetic dataset}, which contains the most important characteristics of the original data and can be openly shared.

\subsection{Original Data}
The original real-world datasets contain records consisting of a series of textual and alphanumerical attributes for \emph{companies}\footnote{Typical attributes for companies are: company names, textual descriptions, headquarter addresses, industry classification codes, market capitalization, \# of employees, etc.} and \emph{securities}\footnote{Typical attributes for securities are: security names, types and unique identifiers based on (inter)national standards such as ISINs (\textbf{I}nternational \textbf{S}ecurities \textbf{I}dentification \textbf{N}umber), CUSIPs (\textbf{C}ommittee on \textbf{U}niform \textbf{S}ecurities \textbf{I}dentification \textbf{P}rocedures), VALORs (Swiss German banking term for a "security"), SEDOLs (\textbf{S}tock \textbf{E}xchange \textbf{D}aily \textbf{O}fficial \textbf{L}ist), etc.} that are traded on certain stock exchanges. Companies can issue multiple securities, but each security belongs exclusively to a single company.

As previously mentioned, these records are continuously updated reflecting real-world events such as bankruptcies, mergers, acquisitions, rebrandings etc. However, updates are not carried out uniformly across all data sources. Consequently, the records across our data sources not only differ due to variations in naming practices (such as using the full company name "Microsoft Corporation" or the stock ticker "MSFT") but also because of the events described above, which are typically absent in other EM benchmarks. In our use case, we have records from around $10$ different data sources, i.e. financial data providers, such as Bloomberg, Reuters, etc.

In order to ease the matching of both company and security records, several international identifier standards have been developed such as International Securities Identification Numbers (ISINs) or Legal Entity Identifiers (LEIs). However, groups of records affected by real-world events are not recognized and/or catalogued by matching approaches relying merely on these identifiers. For example, if a match is made between two companies based exclusively on their LEIs, we might be incorrectly matching an acquirer with its acquiree after an acquisition that led to the overwriting of the LEI of the latter with that of the former. Additionally, many records are missing some or all identifiers and thus can only be matched based on their textual attributes, such as record \#20 in Figure \ref{fig:matching_across_data_sources}, which calls for sophisticated EM methods.

\subsection{Synthetic Benchmark Datasets}
The original datasets are offered by data vendors and distributed under non-disclosure agreements which prevent their open distribution. To overcome this privacy issue, we generate a \emph{synthetic benchmark dataset} from a publicly available set of \rev{1.04M} company records provided by Crunchbase\footnote{See Crunchbase's Basic Export: \href{https://data.crunchbase.com/docs/crunchbase-basic-export}{https://data.crunchbase.com/docs/crunchbase-basic-export}}, which serves as a starting point for recreating our real-world dataset. From the Crunchbase dataset, we extract the \textit{name, city, region, country\textunderscore code, and short\textunderscore description} attributes \rev{of the first 200K records}. In order to create company records that present the same kind of variability as observed among our real-world records, we implement a series of algorithmic modifications of the extracted \rev{records} that we name \textit{data artifacts}. These \textit{data artifacts} modify the original Crunchbase records via rule-based algorithms, much like the data augmentation operators used in pseudo-labeling methods \cite{Li2023,wang2022sudowoodo}. Examples of data artifacts include:

\begin{enumerate}
    \item \textbf{AcronymName (Companies):} Swap a name with its acronym.
    \item \textbf{InsertCorporateTerm (Companies):} Inserts a common term (Inc./Limited/Corp etc.) in all  mentions of a name.
    \item \textbf{CreateCorporateAcquisition/Merger (Companies):} Overwrites records with the attributes of an acquiring investor simulating an acquisition/records the interaction with another company simulating a merger process.
    \item \textbf{ParaphraseAttribute (Companies):} Paraphrase a textual attribute via \rev{the Pegasus summarization} model.
    \item \textbf{MultipleIDs (Securities):} Create new identifiers and assign them to multiple records of a  security.
    \item \textbf{NoIdOverlaps (Securities):} Wipe all overlaps among identifiers of a group of security records.
    \item \textbf{MultipleSecurities (Securities):} Adds new securities of different types such as rights, bonds or units to an issuing company.
\end{enumerate}

\rev{Each of the record groups of the synthetic datasets will have a random combination of \emph{data artifacts} applied to it. Note that multiple \emph{data artifacts} are sequentially applied to each record group and thus their effects become intertwined, generating a big variety of matching challenges across the 200K groups. The size of the generated dataset\footnote{\rev{A maximum of 1.04M (size of the Crunchbase dataset) record groups can be generated. Data artifacts are applied sequentially to each record group thus the complexity of the generation is linear w.r.t the number of record groups.}} and the proportion of record groups to which each \emph{data artifact} is applied to, can be fully parameterized and different choices of generation parameters will lead to datasets with different sizes and proportions of record groups presenting each type of matching challenge. We have calibrated the generation by inspecting the main characteristics of the real dataset\footnote{\rev{We choose generation parameters based on observations made on the subset of manually labeled real records that we use in the experiments. Note that fully characterizing the proportion of matching challenges present in the record groups of the real dataset would require both being able to perfectly match the dataset (i.e. the objective of this work) and then manually annotating the origin of all the challenges each group presents. The latter, however, is an ambiguous task since  combinations of record updates due to real-world events, much like \textit{data artifacts}, often lead to matching challenges whose causes are impossible to discern.}}. Our specific choice of parameters, along with instructions on how to modify them to carry out different generations, is described in the accompanying code\footnote{\rev{See the README file in the folder $\texttt{datainc/datainc\_code}$: \url{https://github.com/FernandoDeMeer/GraLMatch}}}.}

In a real setting, the extent to which different records affected by real-world events should be matched, will vary on a case-by-case basis. Mergers usually involve the creation of a new entity with shares of the original companies being exchanged based on an agreed-upon exchange ratio, which will be different for each of the merging companies depending on their finances. Often, no records are deleted due to a merger, but rather, records of the new merged entity are created. We do \emph{not} consider records involved in simulated mergers as matches. 

On the other hand, acquisitions involve one company absorbing the other, with shareholders of the target company receiving compensation for their shares, which cease to exist along with the absorbed company, whose records are usually deleted in the data sources that record the event. Since the acquiring company has to assume the finances of the acquiree, we consider all records involved in acquisitions as matches.

Records involved in mergers and acquisitions may have some or all of their attributes overwritten due to the event. In Figure \ref{fig:matching_across_data_sources} Records \#10, \#20, \#30 corresponding to "lastminute.com" and \#42 corresponding to "Travix" are involved in a merger and thus \#30 has had some identifiers overwritten with those of \#42 while not being a true match. Records \#11 and \#21 corresponding to "Herotel" and \#33, \#41 corresponding to "Hearst" are involved in an acquisition and thus are all matches. Note that the identifiers of the security record \#21 have been overwritten with those of \#33 and \#44. 

\subsection{Real vs. Synthetic Dataset}\label{subsection:Comparison_syn_vs_real_data}

While far from being able to generate all the variations and interactions present among real records\footnote{Examples of additional difficult phenomena to generate algorithmically while guaranteeing matchability are multiple languages/alphabets, rebrandings, geographical terms, etc.}, \rev{randomly} combining \textit{data artifacts} leads to a \emph{novel, non-trivial dataset for entity group matching}. Below, we describe the main challenges of the dataset:

\begin{enumerate}
    \item Record pairs with \textbf{matching identifiers} are not necessarily true matches since an alignment between textual attributes is needed to detect records affected by \textbf{mergers} or \textbf{acquisitions} events, which may have had some or all of their attributes overwritten.
    \item Matches between records \textbf{lacking matching identifiers} can only be found by matching their textual attributes such as record \#20 of Figure \ref{fig:matching_across_data_sources} which has a different identifier than the one records \#10 and \#30 share and thus has to be matched via its name.
    \item Matches between records whose \textbf{identifiers and textual attributes} are different (e.g. the records of an acquiring company and the acquiree when the data source of the latter has not recorded the acquisition), may only be found by taking into account \emph{transitive information} found through other matched records or be otherwise impossible to detect. For instance, records \#11, \#33 and \#41 of Figure \ref{fig:matching_across_data_sources} can only be matched by matching record \#21 to \#33 or \#41 first.
\end{enumerate}

See Table \ref{tab:dataset_statistics} for a comparison between the generated (synthetic) datasets and their real counterparts.

\begin{table}[]
    \caption{Overview of the general statistics of our datasets. Values marked with a $^\ast$-symbol are estimated, as only a subset of the real datasets were manually checked and labelled.}
    \label{tab:dataset_statistics}
    \resizebox{\linewidth}{!}{
        \begin{tabular}{l|cc|cc}
        \multirow{2}{*}{\textbf{Dataset}}             & \multicolumn{2}{c|}{\textbf{Companies}} & \multicolumn{2}{c}{\textbf{Securities}} \\
                                             & Real            & Synthetic    & Real           & Synthetic     \\ \hline 
        \# of Data Sources                      & $\sim 10$       & 5            & $\sim 10$      & 5             \\ \hline
        \# of Entities                          & $< 200K^\ast$    & 200K         & $< 250K^\ast$   & $\sim 275K$   \\ \hline
        \# of Records                           & $\sim 600K$     & 868K         & $\sim 1M$      & $\sim 984K$   \\ \hline
        \# of Matches                           & $> 1M^\ast$      & $1.5M$       & $> 1.5M^\ast$   & $\sim 1.5M$   \\ \hline
        \begin{tabular}{@{}l@{}}Avg. \# of Matches\\per Entity\end{tabular}           & $7^\ast$         & $7.5$        & $10^\ast$       & $\sim 5.4$    \\ \hline
        \begin{tabular}{@{}l@{}} \% of Records with\\Text Descriptions\end{tabular} & $25 \%$         & $32 \%$      & -              & -            
        \end{tabular}
    }
\end{table}

\section{Entity group matching with GraLMatch}\label{Method}

In this section, we introduce our novel \emph{entity group matching} approach that enables the matching of \rev{record groups} -- as opposed to only matching \rev{record pairs}, which is the common approach of previous Entity Matching research. Note that simply using pairwise matching to solve the entity group matching problem is not sufficient, as we will highlight in this section.

As stated in Section \ref{Section:Introduction}, Entity Matching requires considering the \emph{transitively matched} records of a given \rev{pairwise} matching logic, even if the \rev{pairwise} matching model did not \rev{directly predict each pair of \emph{transitively matched records} to be a match. Including the implied transitive matches of a given set of pairwise match predictions is necessary to obtain each of the expected groups of records (complete graph) of the entities we intend to match.}

\begin{figure}[h!]
    \centering
    \includegraphics[width=0.9\linewidth, keepaspectratio]{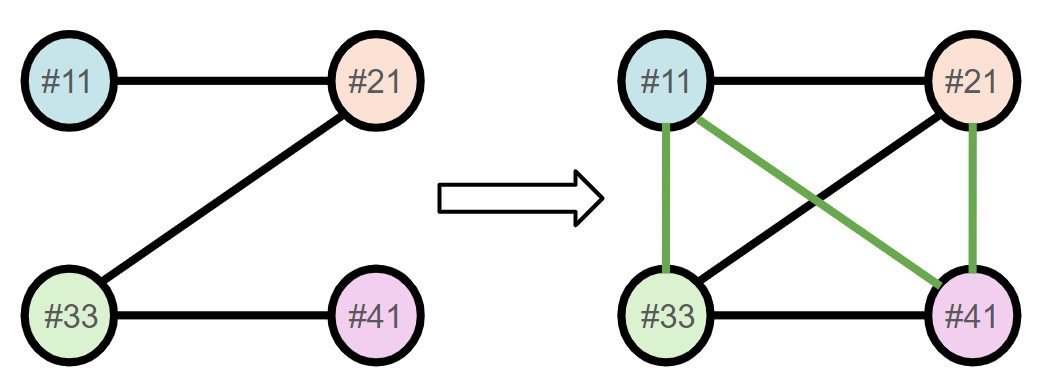}
    \caption{Example of transitive matches between records of Figure \ref{fig:matching_across_data_sources}. On the left side, the pairwise matches ($\#$11 and $\#$21), ($\#$21 and $\#$33) and ($\#$33 and$\#$41) imply the transitive matches of the right side colored in green ($\#$11 and $\#$33), ($\#$11 and $\#$41) and ($\#$21 and $\#$41).}
    \label{fig:transitive_match_example}
\end{figure}

Figure \ref{fig:transitive_match_example} shows an example of a group of records, identified by the \#ID attribute as in Figure \ref{fig:matching_across_data_sources}, with pairwise matches that imply transitive matches. Note that only record $\#$21 contains information about the relationship between all records of the group (the acquisition) and thus the group can only be discovered by matching $\#$21 against either record $\#$33 or $\#$41 and finding all other matches transitively.

\begin{figure*}
  \includegraphics[width=\textwidth, center]{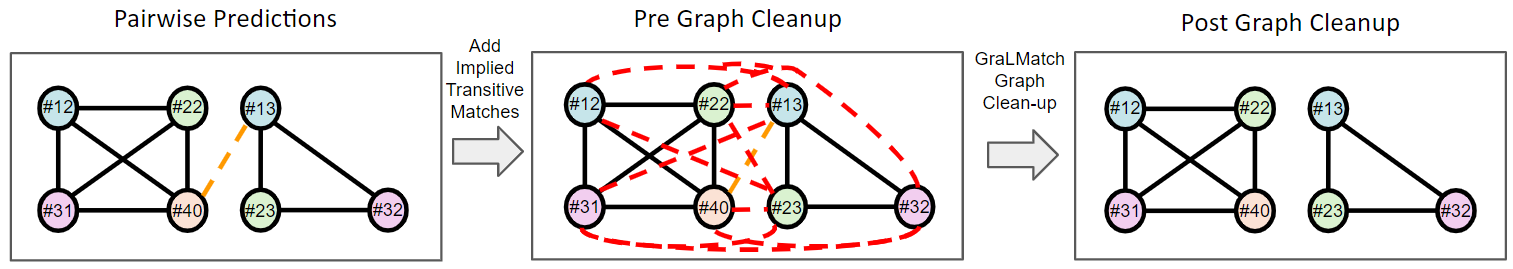}
  \caption{Illustration of entity group matching based on a subset of the records shown in Figure \ref{fig:matching_across_data_sources}. (1) Pairwise predictions: The false positive pairwise match between record \#40 (Crowdstrike) and record \#13 (Crowdstreet)  is illustrated as a dotted orange line. (2) Pre Graph Cleanup: False transitive matches are shown as dotted red lines, e.g. record \#12 (CrowdStrike) is wrongly matched transitively with record \#13 (CrowdStreet). True positive pairwise and final matches are black lines. (3) Post Graph Cleanup: The false pairwise match, originally shown in orange, is eliminated via the GraLMatch Graph Cleanup. The results are two group matches as opposed to one group match resulting from wrong pairwise matching.}
  \label{fig:Sets_of_Matches_considered}
\end{figure*}

\rev{Conversely, Figure \ref{fig:Sets_of_Matches_considered} shows an example of 2 different record groups incorrectly linked through a false pairwise match prediction between records \#40 and \#13. These predictions greatly affect the matching because they lead to all records of both groups being \emph{transitively matched}, which results in numerous false transitive matches.}

Our end-to-end matching methodology is illustrated in Figure \ref{fig:entity group matching Workflow Diagram} and \rev{addresses the transitivity problem through} the following steps: 
\begin{enumerate}
    \item \textbf{Fine-tune Language Models}: We fine-tune a Language Model to perform Sequence Classification in order to predict record pairs as either \texttt{Match} or \texttt{NoMatch}. \rev{We focus on Language Models because they are the state-of-the-art pairwise matching methods.}
    
    \item \textbf{Pairwise matching:} We evaluate a set of record pairs obtained via some blocking\footnote{Blocking is necessary due to the large number of possible candidate pairs and the impossibility of evaluating all due to prohibitive running time.} with the fine-tuned Language Model.
    
    \item \textbf{GraLMatch Graph Cleanup:} Considering the graph $G = (V,E)$ whose nodes and edges represent the records to be matched and the positively predicted matches of the previous step, we delete a series of edges/matches of $G$ with a method that considers the graph implied context (i.e. the connected component it lies on) of each edge/match.

    \item \textbf{Entity Groups:} We output all the connected components of the updated graph as groups of matched records.
\end{enumerate}

\subsection{Pairwise matching with LMs}\label{Entity Matching with LMs}

% \todo[inline]{Note: Language models are usually called Large if they're bigger than GPT2 (1.5 Billion Parameters) source: \url{https://huggingface.co/docs/transformers/model_summary} so we should avoid the term Large Language Model/LLM to refer to the models we use and just say Language Model instead.}

% \begin{itemize}
%     \item Transformer architecture taken from \cite{Brunner2020}, using DistilBert \cite{sanh2019distilbert} for faster execution.
%     \item Sequence encoding based on \cite{Brunner2020}, with special token ideas taken from \cite{Li2023}
% \end{itemize}

Transformer-based Language Models offer several desirable properties for performing pairwise matching over traditionally used heuristics. Namely, through  techniques such as \emph{tokenization, word embeddings, attention mechanisms and pre-training} they are able to jointly process attributes of different data types, either structured or unstructured, making context-aware predictions learned from fine-tuning on large volumes of data.

We employ DistilBERT \cite{sanh2019distilbert} for our experiments, a model obtained via distillation of the pre-trained BERT model. While having $40\%$ fewer parameters than BERT, DistilBERT retains close language understanding capabilities w.r.t. BERT, as evidenced by the similar scores achieved at a series of downstream tasks. Due to the scale of the datasets we aim to match, we choose DistilBERT for faster evaluation speeds given the vast number of comparisons required for the pairwise matching step.

We fine-tune the pre-trained DistilBERT on the binary classification task of predicting record pairs as either \texttt{Match} or \texttt{NoMatch} with a subset of matches and non-matches from the training dataset. In a real setting, only a subset of labeled matches may be available since the labeling effort must be considered. Once the training pairs are decided, we add a final softmax layer to the pre-trained model and then train it for a few epochs as in \cite{Brunner2020, Li2023}.

\subsection{GraLMatch Graph Cleanup}
\label{Method:Graph Cleanup}

Consider the graph $G = (V,E)$ whose nodes and edges are given by the records to be matched and the predicted matching pairs, respectively. The aim of the GraLMatch Graph Cleanup step is to recognize false positive pairwise match predictions to remove them from the final output. The false positives are removed using the \emph{graph implied context} of each pairwise prediction, i.e. the connected component it lies on, or equivalently, all the records \emph{transitively matched} to both records of the pair. Different approaches can be employed to discover good candidate edges for removal, depending on the type of information considered from the connected component.

False positive predictions are often the only link between sets of densely connected nodes \rev{such as the match between records \#40 and \#31 of Figure \ref{fig:Sets_of_Matches_considered}}. The following graph-based methods allow us to filter such edges and mark them for removal:

\begin{enumerate}
    \item \textit{Minimum Edge Cut:} Given a graph $G= (V,E)$ a \emph{minimum edge cut} $S$ is a set of edges $S \subseteq E$ of minimum cardinality (i.e. $|V_S|$ if $S = (V_S, E_S)$) that disconnects G.
    \item \textit{Edge Betweenness Centrality:} Given a graph $G= (V,E)$, the Betweenness Centrality of an edge $e \in E$
    is the sum of the fraction of all-pairs shortest paths that pass through $e$:
    $$
    c_B(e)=\sum_{s, t \in V} \frac{\sigma(s, t \mid e)}{\sigma(s, t)}
    $$
    where $\sigma(s, t)$ is the number of shortest-paths between nodes $s$ and $t$, and $\sigma(s, t \mid e)$ is the number of those paths passing through edge $e$.

\end{enumerate}

The matching of large datasets may lead to numerous large connected components, each potentially containing multiple good candidate edges for removal. \rev{Note that removing the edges of a connected component belonging to a \textit{Minimum Edge Cut} guarantees disconnecting said component, while the same is not true when removing the edges with highest \textit{Edge Betweennes centrality}. Both techniques have a time complexity of $O(mn)$  where $n$ is the number of records and $m$ the number of edges of a given connected component \cite{brandes2001faster, esfahanian2013connectivity}. However, in practice, the \emph{Minimum Edge Cut} tends to have a lower run-time, even if both worst-case time complexities are identical. We generally expect the \textit{Edge Betweennes centrality} to remove less true positive edges while being slower than the \emph{Minimum Edge Cut}.} 

\rev{Algorithm \ref{alg:Graph Cleanup} shows how the two techniques described above are used by our GraLMatch Graph Cleanup. Size threshold $\gamma$ specifies which of the techniques should be used, depending on the size of the connected component to clean up, whereas $\mu$ specifies the desired maximum size of all produced record groups.}

\rev{We set size threshold $\mu$ to be equal to the number of data sources. This is ideal in settings where each group is expected to have at most one record per data source, as is the case in our datasets of interest. However, in settings where there is not a specified number of data sources and/or we expect a lot of record groups of different sizes, as in the case of web-scrapped records, Algorithm \ref{alg:Graph Cleanup} will not be ideal and other Graph Cleanup methods able to produce groups of heterogeneous sizes should be considered.}

\begin{algorithm}
\caption{GraLMatch Graph Cleanup}
\label{alg:Graph Cleanup}
\hspace*{\algorithmicindent} \textbf{Input:} matches graph $G = (V,E)$, size thresholds $\gamma$ and $\mu$\\
\hspace*{\algorithmicindent} \textbf{Output:} List of connected components of $G$ after cleanup. 
\begin{algorithmic}[1]
\State $C = \{c_1, c_2, ..., c_n\}$ connected components of $G$
\State $c^* \gets \argmax(\{\,|c_i| \mid c_i \in C\})$

\While{$ |c^*| > \gamma$}:
\State  $E_{\text{mincut}} \gets \text{MinEdgeCut($c^*$)}$
\State  $G \gets (V, E \setminus E_{\text{mincut}})$
\State $c^* \gets \argmax(\{\,|c_i| \mid c_i \in C\})$
\EndWhile
\While{$ |c^*| > \mu$}:
\State  $e_{\text{maxBC}} \gets \argmax(\{BtwCent(e) \mid e \in E^* \})$
\State  $G \gets (V, E \setminus e_{\text{maxBC}})$
\State $c^* \gets \argmax(\{\,|c_i| \mid c_i \in C\})$

\EndWhile
\State \textbf{Output:} $C = \{ \text{c}_1, \text{c}_2, \ldots, \text{c}_n \}$
\end{algorithmic}
\end{algorithm}

\begin{comment}
Alternatively, considering the attributes of the neighbouring nodes of each edge, the discovery of candidate edges could be also framed as 
an ML problem. This approach is justified by the fact that the context characteristics of true $\&$ false positive matches vary for each dataset. A given dataset may for example have a lot of record groups that can only be fully matched via a single match that connects two sets of densely connected nodes (see Figures \ref{fig:matching_across_data_sources} and \ref{fig:transitive_match_example} for an example of such record ), which would be likely flagged as a false positive by the graph heuristics previously described.

Aspects such as which neighbouring nodes' attributes to consider, the encoding method, the gathering of training examples (especially including expected edge cases) with its associated labelling effort, the type of LM to be used for prediction and its evaluation speed are all important considerations given the difficulty and scale of the task (since it consists on identifying the mistakes made by another LM). We leave this line of research for future work.
\end{comment}

\subsubsection{Pre Graph Cleanup}

Some sets of pairwise match predictions lead to exceedingly large connected components in $G$. In order to avoid long running times of Algorithm \ref{alg:Graph Cleanup}\footnote{Both of the edge removal techniques used during the GraLMatch Graph Cleanup remove a few edges at a time and are thus slow at cleaning up exceedingly large connected components.}, we further apply the following pre-cleanup technique:

%\begin{itemize}
    \textit{\textbf{Company datasets:}} We remove all positively predicted matches obtained through the \textit{Token Overlap} blocking in connected components larger than 50 records. See Section \ref{Experiments:Blockings} for details.
%\end{itemize}

\section{Experiments}\label{Experiments}

In this section, we describe our experimental setup\rev{\footnote{\rev{Our code can be found here: \url{https://github.com/FernandoDeMeer/GraLMatch}}}}. The goal is to address the following research questions:

\begin{itemize}
\item What is the \emph{performance} of state-of-the-art Entity Matching algorithms for \emph{pairwise and entity group matching} applied to our challenging datasets?
\item Can our \emph{novel algorithm GraLMatch boost the performance existing algorithms for entity group matching}? 
\end{itemize}

\rev{All the experiments run in this paper were conducted on an Ubuntu machine with a Nvidia Tesla T4 GPU, 16 VCPUs, and 64 GB of RAM.}

\subsection{Datasets}\label{Experiments:Datasets}

We now describe both the real and synthetic datasets used for our experiments. Our aim is to \textit{match multiple real-world datasets} made up of records of companies and financial securities. As described in Section \ref{Section: Syn Dataset Generation}, due to the confidential nature of this data, we have produced a dataset of groups of records from publicly available data that present similar matching challenges as their real counterparts.

\subsubsection{Real Companies and Securities Datasets}

In the experiments, we use a small subset of 63.5k human-labeled security record groups obtained through matching identifier codes combined with 1.5k manually found edge case record groups (with missing identifiers, multiple identifiers for matching securities, mergers, acquisitions etc.), totaling to 65K companies and securities from 8 different data sources. For the company records, we use the corresponding issuers of each of the securities. \rev{These datasets are small in relation to the entire real dataset and contain a very low proportion of challenging record groups, due to the difficulty involved in manually finding them, see Table \ref{tab:dataset_statistics} for reference.}

\subsubsection{Synthetic Companies and Securities Datasets}
We perform experiments on the synthetically generated datasets described in Section \ref{Section: Syn Dataset Generation} in order to estimate the different models' performance at entity group matching in a real world setting. \rev{These datasets are closer in scale to the entire real datasets than the human-labeled set of real records described above, see Table \ref{tab:dataset_statistics} for reference. Consequently, the results we obtain in these datasets will give us a better approximation to our performance in the real use case which involves the entire real dataset.} We make these datasets, as well as the code to generate them, fully accessible to the wider research community for utilization.

\subsubsection{Train, Validation and Test Splits} \label{Experiments:Splits}

In order to fine-tune and evaluate each machine learning model, we divide the records of the datasets above into \textit{train, validation and test} splits, each containing all the records belonging to $60\%/20\%/20\%$ of the ground truth record groups\footnote{The percentages roughly correspond to the $\%$ of records in each split but there are small variations because record groups vary in size.}. \rev{We fine-tuned models with all the positive pairs of each split} and add randomly sampled negative pairs with a ratio of $5:1$ negative pairs for each positive one. We split along the record groups to make sure that the set of true matches of each entity belongs exclusively to one split, preventing models from memorizing pairs. 

\subsubsection{\rev{WDC Products}}
\rev{Additionally, we evaluate our matching pipeline on the WDC Products benchmark dataset \cite{Peeters2024WDCPA}. More specifically, we experiment on the large dataset variant with $80\%$ corner cases and a test set with $100\%$ unseen entities.}

\subsection{Machine Learning Models}

For our experiments we use Ditto \cite{Li2023}, the state-of-the-art machine learning model for Entity Matching, as our baseline. We also considered employing a large language model, LlaMa2 7B, by framing the classification task as a text generation instance via prompt-engineering and recovering answers via a regular expression. Initial experiments showed that LlaMa2 took on average 7 seconds to generate an answer per candidate pair which leads to exceedingly long running times for the pairwise matching step (90+ days). We expect this also to be the case with comparably sized and bigger large language models. \rev{We use the following specifications for each model}:

\begin{itemize}
    \item \textbf{DITTO (128)}: In accordance with our focus on speed, this DITTO \cite{Li2023} variant uses the DistilBERT \cite{sanh2019distilbert} model due to the size of our datasets, similar to the choice in the original DITTO paper for their larger datasets. Additionally, the DITTO (128) variant uses a max sequence length of 128 tokens, half the size used in the original paper. Please note, that DITTO uses a different encoding scheme which wraps column names and values with special start and end tokens. For example, the value \texttt{Zurich} in the \texttt{city} column would be encoded as \texttt{[col] city [val] Zurich}. This increases the amount of tokens required to encode the same value information, but adds more structure and the information of column names.
    
    \item \textbf{DITTO (256)}: This variant of DITTO is identical to the DITTO (128) variant, except that it uses sequences with up to 256 tokens (as in the original paper).
    
    \item \textbf{DistilBERT (128)-ALL}:  We fine-tune DistilBERT with a maximum token sequence length of 128  on all of the pairs in the train splits.
\end{itemize}

\subsubsection{\rev{Sensitivity Analysis}}

\rev{
In order to study the sensitivity of our pipeline w.r.t the amount of available labeled data, we fine-tune DistilBERT (128) on small sets of training pairs from the synthetic companies and securities datasets that could be acquired with a moderate manual labeling effort:}
\begin{itemize}

    \item \textbf{DistilBERT (128)-15K}:  We fine-tune DistilBERT with a maximum token sequence length of 128 on a set of pairs obtained by \emph{filtering the first 10K/5K pairs from the train/val splits}. We discard those whose records have been involved in an \emph{acquisition} or cannot all be matched via identifier overlaps. 
\end{itemize}

\rev{Additionally, in order to study the sensitivity w.r.t the size thresholds $\gamma$ and $\mu$, we run the matching pipeline with \textbf{DistilBERT (128)-ALL} and the following modifications:} 

\begin{itemize}
    \item \rev{\textbf{DistilBERT (128)-ALL-MEC}: We run the Graph Cleanup with $\gamma = \mu$, that is, we only use the \textit{Minimum Edge Cut (MEC)} to remove edges.}
    \item \rev{\textbf{DistilBERT (128)-ALL ($\frac{1}{2}\gamma$)}: We set the size threshold $\gamma$ to half of its value in Table \ref{tab:blockings_datasets}  (rounded down).} 
    \item \rev{\textbf{DistilBERT (128)-ALL-BC}: We run the Graph Cleanup with  $\gamma = \infty$, that is, we only use the \textit{Betweenness Centrality (BC)} to remove edges.}

\end{itemize}

\rev{ We run the 3 previous variants on the Synthetic Companies dataset, since it is the dataset with the biggest number of candidate pairs (and thus connected components), see Table \ref{tab:blockings_datasets}. For all models, we fine-tune for 5 epochs and select the epoch with the lowest validation loss.}

\subsection{End-to-End entity group matching Experiments} \label{Experiments:Matching}
As mentioned previously, our aim is to \emph{match multiple real-world datasets} for which a ground-truth is initially unavailable. To estimate our performance on this challenging task, we carry out the end-to-end entity group matching process on the datasets described in Section \ref{Experiments:Datasets}.
\subsubsection{Blockings}
\label{Experiments:Blockings}
In order to start the entity group matching process, we first need to obtain a subset of candidate record pairs through a combination of blockings in order the reduce the complexity and running time of the matching process. We employ the following blockings:
\begin{enumerate}
    \item \textit{\textbf{ID Overlap:}} Finds candidate pairs based exclusively on the overlap of identifier attributes. Typical examples of IDs for \emph{securities} are ISINs, CUSIPs, VALORs or SEDOLs as shown in the bottom part of Figure \ref{fig:matching_across_data_sources}. In the case of \emph{company} records, we evaluate against the companies whose associated securities have a matching identifier with any of the securities issued by each company record. This blocking is equivalent to the benchmark heuristic often used to match these types of financial records. It leads to few candidate pairs containing positive and negative pairs as described in Section \ref{subsection:Comparison_syn_vs_real_data}.
    \item \textit{\textbf{Token Overlap:}} Considers each record as the list of tokens resulting from its tokenization and selects as candidate pairs those involving the record and the top $n$ records with most overlapping tokens across different data sources. This blocking aims to find good candidate pairs for matches based on text alignment.
    \item \textit{\textbf{Issuer Match (Securities Only):}} For each security record, consider as candidate pairs those involving all other securities issued by companies previously matched to the security's issuer. This blocking allows to match pairs of securities with non-matching identifiers and generic names based on a previous matching of their issuers.
\end{enumerate}

We combine the blockings described above differently for each dataset as detailed in Table \ref{tab:blockings_datasets}, producing different sets of candidate pairs. We predict each of the candidate pairs as either \texttt{Match} or \texttt{NoMatch}, leading to a set of pairwise match predictions for each model/dataset pair.

\begin{table}[H]
    \centering
    \caption{Blockings applied, number of records, candidate pairs and size thresholds in the entity group matching experiment for each dataset.}

    \begin{tabular}{c| ccccc}
         & Blockings& \begin{tabular}{@{}c@{}}$\#$ of \\ Records\end{tabular}
           & \begin{tabular}{@{}c@{}}$\#$ of \\ Candidate \\Pairs\end{tabular} & $\gamma$ & $\mu$\\\toprule
    \begin{tabular}{@{}c@{}}\textit{Real} \\ \textit{Companies}\end{tabular}    & \begin{tabular}{@{}c@{}}\textit{ID Overlap} \\ \textit{Token Overlap}\end{tabular} & 6.3K & 51K & 40 & 8\\\midrule
    \begin{tabular}{@{}c@{}}\textit{Synthetic} \\ \textit{Companies}\end{tabular}    & \begin{tabular}{@{}c@{}}\textit{ID Overlap} \\ \textit{Token Overlap}\end{tabular} & 174K & 1.14M & 25 & 5\\\midrule

    \begin{tabular}{@{}c@{}}\textit{Real} \\ \textit{Securities}\end{tabular}  & \begin{tabular}{@{}c@{}}\textit{ID Overlap} \\ \textit{Issuer Match}\end{tabular} & 12.8K & 41K & 40 & 8 \\\midrule
    \begin{tabular}{@{}c@{}}\textit{Synthetic} \\ \textit{Securities}\end{tabular} & \begin{tabular}{@{}c@{}}\textit{ID Overlap} \\ \textit{Issuer Match}\end{tabular} & 197K & 826K & 25 & 5 \\\midrule

    \begin{tabular}{@{}c@{}} \textit{\rev{WDC}} \\ \textit{\rev{Products}}\end{tabular}  & \rev{Token Overlap} & \rev{1K} & \rev{9.1K} & \rev{25} & \rev{5}\\\midrule

    \end{tabular}
    \label{tab:blockings_datasets}
\end{table}

\subsubsection{Impact of GraLMatch Graph Cleanup}\label{subsubsec:Impact of Graph Cleanup}

In order to illustrate the importance of the GraLMatch Graph Cleanup step with respect to the final entity group matching, we calculate \emph{precision, recall and F1} at three different stages of our end-to-end matching pipeline:

\begin{enumerate}
    \item \textit{\textbf{Stage 1: Pairwise matching:}} The \emph{positively predicted pairs} from all of the candidate pairs produced by the combination of blockings in each dataset.
    \item \textit{\textbf{Stage 2: Pre Graph Cleanup:}} Considering the graph $G=(V,E)$ whose nodes and edges represent the records to be matched and the pairwise match predictions respectively, we incorporate all of the edges missing from each connected component to make the component a \textit{complete} subgraph. That is, we add all the edges connecting pairs of \textit{transitively matched} records by the pairwise match predictions.
    \item \textit{\textbf{Stage 3: Post Graph Cleanup:}} We run the GraLMatch Graph Cleanup on the graph $G=(V,E)$ (with the pairwise match predictions only) and then add all of the transitive matches to the list of cleaned up connected components produced by the GraLMatch Graph Cleanup.
\end{enumerate}

Note that the scores achieved for \emph{pairwise matching} are not equivalent to those obtained during fine-tuning. The recall is expected to be lower since some \emph{true positives are discarded by the blocking}, whereas all the true positives are available and evaluated during fine-tuning. Additionally, the precision is also expected to be lower since the set of candidate pairs contains more challenging negative pairs than the ones added randomly during fine-tuning. 

Considering all of the matches included in the \emph{Pre Graph Cleanup} set is justified because they represent the group assignment implied by the \emph{pairwise matching}. We will show how usually the Pre Graph Cleanup scores are considerably low, since very few false positive predictions can lead to a large amount of false transitive matches, especially when large connected components are produced by a set of pairwise match predictions. 

The \emph{Post Graph Cleanup} scores represent the final group assignment of our methodology. See Figure \ref{fig:Sets_of_Matches_considered} for an illustration of all three sets of matches and the phenomenon described above.

It is important to note that the scores achieved with pairwise matching should not be compared to Pre and Post Graph Cleanup scores since they do not represent a group assignment (they ignore the \textit{transitively matched} records) but rather only represent a model's pairwise matching performance.

\subsubsection{Graph Metric}
Along with precision, recall and F1 scores, we calculate the following graph-based metric:

\begin{itemize}
    \item \textbf{Cluster Purity Score:} Given an entity group matching $M$ composed of complete subgraphs $\{c_i = (V_i, E_i)\}_{i=1}^N$ its cluster purity is calculated as: 
    \begin{center}
        $\text{ClPur}  = \frac{1}{\sum_{i=1}^N |V_i|}\sum_{i=1}^N |V_i|\frac{c_{TP,i}}{|E_i|}$
    \end{center}
    where $c_{TP,i}$ is the number of true positive matches in subgraph $c_i$.
    
\end{itemize}

The Cluster Purity Score is thus the average of the number of correct matches per record group weighted by the size of each group. It indicates how reliable downstream tasks based on the aggregation of calculations made on record groups will be.

\section{Results}\label{Results}

% In this section we will discuss the results of the fine-tuning and entity group matching experiments.

\begin{table*}[]
\caption{Overview of the scores achieved by fine-tuned models on test pairs.}
\label{tab:test_pairwise_table}
\begin{tabular}{ll|rrr|r}
                          &                         & \multicolumn{3}{c|}{\textbf{Pairwise Matching Performance}}                           & \multicolumn{1}{l}{}                         \\
\textbf{Dataset} & \textbf{Model} & \textbf{Precision} & \textbf{Recall} & \textbf{F1 Score} & \textbf{Training Time} \\ \hline
Real Companies            & DITTO (128) & 68.82 $\pm$ 0.00              & 83.49 $\pm$ 0.00             & 75.11 $\pm$ 0.00              &  \textbf{18.74 h}           \\
                          & DITTO (256) & 99.90 $\pm$ 0.09              & 99.67 $\pm$ 0.13             & \textbf{99.78 $\pm$ 0.11}              & 33.59 h            \\
                          & DistilBERT (128)-ALL&       99.93 $\pm$ 0.02             &    99.56 $\pm$  0.04               &     99.73 $\pm$ 0.02            &                    23.25 h\\ \hline
Synthetic Companies       & DITTO (128)        & 99.45 $\pm$ 0.06              & 96.70 $\pm$ 0.30             & 98.15 $\pm$ 0.01              & 85.11 h            \\
                          & DITTO (256)        & 99.55 $\pm$ 0.01              & 96.88 $\pm$ 0.01             & \textbf{98.20 $\pm$ 0.01}              & 86.39 h            \\
                          & DistilBERT (128)-15K     &        99.35 $\pm$ 0.03           &   94.77 $\pm$ 2.74                &     96.99 $\pm$ 1.41               & \textbf{11.32 h}\\
                          & DistilBERT (128)-ALL     &     99.28 $\pm$ 0.13                &    96.09  $\pm$  0.06               &    97.66 $\pm$ 0.03                & 93.28 h          \\ \hline
Real Securities           & DITTO (128)        & 25.55 $\pm$ 7.40              & 69.00 $\pm$ 43.84            & 33.89 $\pm$ 0.17              & 22.71 h            \\
                          & DITTO (256)        & 99.94 $\pm$ 0.01              & 99.13 $\pm$ 0.02             & \textbf{99.53 $\pm$ 0.01}              & 37.88 h            \\
                          & DistilBERT (128)-ALL          & 99.48 $\pm$ 0.18              & 99.48 $\pm$ 0.14            & 99.47 $\pm$ 0.02              & \textbf{20.96 h}            \\ \hline 

\hline
Synthetic Securities      & DITTO (128)        & 57.82 $\pm$ 2.12              & 56.00 $\pm$ 11.64             & 56.47 $\pm$ 5.01              & 94.43 h           \\
                          & DITTO (256)        & 85.51 $\pm$ 0.55              & 91.35 $\pm$ 0.51             & \textbf{88.33 $\pm$ 0.06}              & 122.44 h            \\
                          & DistilBERT (128)-15K          & 94.03 $\pm$ 0.41                 &  61.11 $\pm$  0.33              &                  73.26 $\pm$ 0.24  &  \textbf{11.62h}                   \\
                          & DistilBERT (128)-ALL          & 90.96 $\pm$ 0.77               & 70.55 $\pm$ 0.52                 & 79.46 $\pm$ 0.06          & 103.99 h                   \\ \hline 
\rev{WDC Products}  & \rev{DITTO (128)}  & \rev{35.92 $\pm$ 0.01}  & \rev{63.20 $\pm$ 2.83} & \rev{45.81 $\pm$ 1.56}  & \rev{27.63 min} \\
& \rev{DITTO (256)}  & \rev{48.45 $\pm$ 5.39}  & \rev{72.30 $\pm$ 7.21}  & \rev{\textbf{57.71 $\pm$ 1.53}}  & \rev{40.28 min} \\
& \rev{DistilBERT (128)-ALL}         &                 \rev{46.24} $\pm$ \rev{0.96}   & \rev{76.33} $\pm$ \rev{1.70}  & \rev{57.58} $\pm$ \rev{0.96}                   & \rev{\textbf{26.79 min}}             \\\hline
\end{tabular}

\end{table*}

\subsection{Fine-Tuning}
Table \ref{tab:test_pairwise_table} shows the precision, recall and F1 scores of  each fine-tuned model on test split pairs. 

On the companies datasets, both real and synthetic, all models reach close-to-perfect scores with the exception of DITTO (128) on real companies, which we speculate misses important elements of some inputs due to its encoding method. DistilBERT(128)-ALL reaches a very similar performance to that of DITTO (256) while using half of the input tokens. DistilBERT(128)-15K achieves a lower recall than all other setups, which is to be expected since it is trained with much fewer samples, yet reaches a competitive F1 score while taking 1/6th of the time to train than the other setups.

On the securities datasets, we similarly observe that DITTO (128) struggles to perform due to its encoding method, this time likely due to the identifier attributes which lead to long sequences of uninformative tokens. We observe that on real securities DistilBERT(128) and DITTO (256) reach a similar F1 score while on the synthetic securities, DITTO (256) reaches the overall best F1 score but struggles with precision. This is likely due to the fact that matches between records whose issuers have been involved in a \emph{data drift} event do not present matching identifiers and have generic names (see for example the securities records \#31 and \#40 of Figure \ref{fig:matching_across_data_sources}, both belonging to records of the entity "Crowdstrike") but DITTO (256)'s extensive encoding method seems to allow it to capture some of these challenging pairs. DITTO (256) seems more likely to make positive predictions, which captures some \emph{data drift} pairs but also false positives.

\rev{On the WDC Products dataset DistilBERT(128)-ALL achieves a slightly worse performance to that of RoBERTa as reported in \cite{Peeters2024WDCPA}. The result is expected due to the training optimizations of RoBERTa and its larger model size compared to DistilBERT(128). }

\rev{DistilBERT(128)-ALL also matches the performance of DITTO (256) while using half of the input tokens and outperforms DITTO (128), which indicates DITTO's encoding and training optimizations do not translate into a better performance in this dataset for the smaller model setup.}
% Some of the scores achieved are close to perfect and might make it seem that the datasets are trivial to match. However, in the following section we show how these scores do not guarantee a good entity group matching.
% Pairwise results on test splits

% GROUP MATCHING

\begin{table*}[]

\caption{Overview of the scores achieved in the \textit{entity group matching \rev{with Blocking and GraLMatch}. To achieve the Post Graph Cleanup scores} our novel algorithm GraLMatch is applied.}
\label{tab:group_matching_table}
\resizebox{\textwidth}{!}{%
\begin{tabular}{ll|rrr|rrrr|rrrr|r}
                          &                         & \multicolumn{3}{c|}{\begin{tabular}{@{}c@{}}\textbf{Pairwise Matching Performance} \\ (pairs from blocking)\end{tabular}}                         &\multicolumn{8}{c|}{\begin{tabular}{@{}c@{}}\textbf{Entity Group Matching Performance} \\ (including implied transitive matches)\end{tabular}}             & \multicolumn{1}{c}{}             \\\\
 & & & & & \multicolumn{4}{c|}{\textbf{Pre Graph Cleanup}} & \multicolumn{4}{c|}{\textbf{Post Graph Cleanup}}&\\
\textbf{Dataset} & \textbf{Model} & \textbf{Precision} & \textbf{Recall} & \textbf{F1 Score} &  \textbf{Precision} & \textbf{Recall} & \textbf{F1 Score} & \textbf{Cluster Purity}     &\textbf{Precision} & \textbf{Recall} & \textbf{F1 Score} & \textbf{Cluster Purity}     & \textbf{Inference Time} \\ \hline
  \begin{tabular}{@{}c@{}}Real \\ Companies\end{tabular}           & DITTO (128)         & 23.66 $\pm$ 0.00  & 99.64 $\pm$ 0.00  & 38.24 $\pm$ 0.00  & 0.05 $\pm$ 0.00  & 99.66 $\pm$ 0.02  & 0.10 $\pm$ 0.00  & 0.00 $\pm$ 0.00  & 99.86 $\pm$ 0.12  & 98.23 $\pm$ 0.55  & \textbf{99.06 $\pm$ 0.34}  & 1.00 $\pm$ 0.00 & 6.7 min  \\
                          & DITTO (256)         & 23.66 $\pm$ 0.00  & 99.64 $\pm$ 0.00  & 38.24 $\pm$ 0.00  & 23.52 $\pm$ 0.01  & 99.68 $\pm$ 0.04  & 38.06 $\pm$ 0.02  & 0.00 $\pm$ 0.00  & 98.42 $\pm$ 0.01  & 99.70 $\pm$0.02  & 99.05 $\pm$ 0.01  & 0.99 $\pm$ 0.00 & 6.6 min  \\
                          &  DistilBERT (128)-ALL          & 94.06 $\pm$ 4.83         & 99.27 $\pm$ 0.24       &  96.53 $\pm$ 2.49        &  49.07 $\pm$ 34.11  &  99.73 $\pm$ 0.16  & 56.92 $\pm$ 38.6  &  0.80 $\pm$ 0.16  & 86.90 $\pm$ 5.07   & 96.98 $\pm$ 3.69  & 91.64 $\pm$ 4.25  & 0.93 $\pm$ 2.50  & \textbf{3.5min}         \\
                          \hline

\begin{tabular}{@{}c@{}}Synthetic \\ Companies\end{tabular}        & DITTO (128)       & 33.16 $\pm$ 0.00  & 81.73 $\pm$ 0.00  & 47.18 $\pm$ 0.00  & 0.00 $\pm$ 0.00  & 83.06 $\pm$ 0.28  & 0.00 $\pm$ 0.00  & 0.00 $\pm$ 0.00 & 99.09 $\pm$ 0.13 & 36.94 $\pm$ 2.84 & 53.78 $\pm$ 2.99  & 0.99 $\pm$ 0.00  &    1h 26min         \\
& DITTO (256)        & 33.16 $\pm$ 0.00  & 81.73 $\pm$ 0.00  & 47.18 $\pm$ 0.00  & 0.00 $\pm$ 0.00  & 83.66 $\pm$ 0.57  & 0.00 $\pm$ 0.00  & 0.00 $\pm$ 0.00 & 99.07 $\pm$ 0.30  & 38.06 $\pm$ 3.90  & 54.93 $\pm$ 4.02  & 0.99 $\pm$ 0.00  & 1h 20min \\
& DistilBERT (128)-15K     &   83.08 $\pm$  4.54               &  77.48 $\pm$ 0.58   &  80.11 $\pm$ 1.92            & 0.01 $\pm$ 0.01& 82.31 $\pm$ 1.44 & 0.02 $\pm$ 0.02 & 0.42 $\pm$  0.12 & 98.06 $\pm$ 0.43 & 57.90 $\pm$ 9.17  & \textbf{72.34 $\pm$ 7.47}    & 0.98 $\pm$ 0.00       & \textbf{1h 15min}          \\
& DistilBERT (128)-ALL    &   77.03 $\pm$ 3.83                &   79.46 $\pm$ 0.05               &    78.18 $\pm$ 2.00               & 0.00 $\pm$ 0.00 & 82.26 $\pm$ 0.78  & 0.00 $\pm$ 0.00 & 0.23 $\pm$ 0.05& 98.76 $\pm$ 0.26 & 43.31 $\pm$ 5.10 & 60.03 $\pm$ 5.00 &   0.99 $\pm$ 0.00         &    1h 15min        \\
& \rev{DistilBERT (128)-ALL-MEC} & \rev{77.03 $\pm$ 3.83} & \rev{79.46 $\pm$ 0.05} & \rev{78.18 $\pm$ 2.00} & \rev{0.00 $\pm$ 0.00} & \rev{82.26 $\pm$ 0.78}  & \rev{0.00 $\pm$ 0.00} & \rev{0.23 $\pm$ 0.05} & \rev{98.57 $\pm$ 0.28} &  \rev{42.79 $\pm$ 4.91} & \rev{59.50 $\pm$ 4.83}  & \rev{0.99 $\pm$ 0.00}  & \rev{1h 14min} \\ 
& \rev{DistilBERT (128)-ALL ($\frac{1}{2}\gamma$)} & \rev{77.03 $\pm$ 3.83} & \rev{79.46 $\pm$ 0.05} & \rev{78.18 $\pm$ 2.00} & \rev{0.00 $\pm$ 0.00} & \rev{82.26 $\pm$ 0.78}  & \rev{0.00 $\pm$ 0.00} & \rev{0.23 $\pm$ 0.05} & \rev{98.79 $\pm$ 0.25}  & \rev{43.23 $\pm$ 5.08}  & \rev{59.96 $\pm$ 4.97}  & \rev{0.99 $\pm$ 0.00} & \rev{1h 15min} \\
& \rev{DistilBERT (128)-ALL-BC} &   \rev{77.03 $\pm$ 3.83}                &   \rev{79.46 $\pm$ 0.05}               &    \rev{78.18 $\pm$ 2.00}               & \rev{0.00 $\pm$ 0.00} & \rev{82.26 $\pm$ 0.78}  & \rev{0.00 $\pm$ 0.00} & \rev{0.23 $\pm$ 0.05} & \rev{98.76 $\pm$ 0.26} & \rev{43.31 $\pm$ 5.10} & \rev{60.03 $\pm$ 5.00} &   \rev{0.99 $\pm$ 0.00}         &    \rev{1h 17min}\\\hline
\begin{tabular}{@{}c@{}}Real \\ Securities\end{tabular}            & DITTO (128)        & 19.96 $\pm$ 0.00  & 91.99 $\pm$ 0.00  & 32.80 $\pm$ 0.00  & 19.95 $\pm$ 0.01  & 92.10 $\pm$ 0.02  & 32.80 $\pm$ 0.02  & 0.20 $\pm$ 0.00 & 19.35 $\pm$ 0.46 &  17.59 $\pm$ 4.77 &  18.28 $\pm$ 2.84 &  0.19 $\pm$ 0.00 & 4.8 min       \\
& DITTO (256)        & 19.96 $\pm$ 0.00  & 91.99 $\pm$ 0.00  & 32.80 $\pm$ 0.00  & 19.94 $\pm$ 0.00  & 92.11 $\pm$ 0.00  & 32.78 $\pm$ 0.00  & 0.20 $\pm$ 0.00 & 19.70 $\pm$ 0.01 &  20.93 $\pm$ 0.00 &  20.30 $\pm$ 0.01 &  0.19 $\pm$ 0.00 & 4.5 min       \\
& DistilBERT (128)-ALL           & 99.76 $\pm$ 0.03         & 97.77 $\pm$ 0.01       &  98.76 $\pm$ 0.01       & 99.73 $\pm$ 0.02  & 98.08 $\pm$ 0.04 & 98.90 $\pm$ 0.01 & 1.00 $\pm$ 0.00 & 99.73 $\pm$ 0.02 & 98.00 $\pm$ 0.04 & \textbf{98.86 $\pm$ 0.01}  & 1.00 $\pm$ 0.00 & \textbf{2.6 min}            \\\hline 

\begin{tabular}{@{}c@{}}Synthetic \\ Securities\end{tabular}       & DITTO (128)        & 97.26 $\pm$ 0.00  & 52.51 $\pm$ 0.00  & 68.20 $\pm$ 0.00  & 96.39 $\pm$ 0.10  & 54.58 $\pm$ 0.52  & 69.69 $\pm$ 0.45  & 0.98 $\pm$ 0.00  & 98.22 $\pm$ 0.22  & 44.88 $\pm$ 4.88  & 61.54 $\pm$ 4.65  & 0.99 $\pm$ 0.01 & 29.6 min  \\
& DITTO (256)        & 97.26 $\pm$ 0.00  & 52.51 $\pm$ 0.00  & 68.20 $\pm$ 0.00  & 96.23 $\pm$ 0.25  & 57.08 $\pm$ 0.00  & 71.66 $\pm$ 0.06  & 0.98 $\pm$ 0.00  & 98.31 $\pm$ 0.33  & 56.68 $\pm$ 0.12  & 71.90 $\pm$ 0.18  & 0.99 $\pm$ 0.01 & 29.0 min  \\ 
&  DistilBERT (128)-15K  & 97.26 $\pm$ 0.00 & 57.06 $\pm$ 0.03 & 71.59 $\pm$ 0.04 & 96.05 $\pm$ 0.00 & 57.06 $\pm$ 0.03 & 71.59 $\pm$ 0.02 & 0.98 $\pm$ 0.00 & 98.08 $\pm$ 0.00 & 56.56 $\pm$ 0.04 & 71.71 $\pm$ 0.03 &  0.98 $\pm$ 0.00 & \textbf{23.3 min}  \\ & DistilBERT (128)-ALL         &  95.58 $\pm$ 2.03 & 53.28 $\pm$ 0.88 & 68.40 $\pm$ 0.20 & 87.81 $\pm$ 9.73 & 58.40 $\pm$ 1.50 & 69.82 $\pm$ 2.27 & 0.94 $\pm$ 0.04 & 96.70 $\pm$ 1.60 &  57.52 $\pm$ 1.02& \textbf{72.11 $\pm$ 0.34}& 0.97 $\pm$ 0.01 &  23.4 min           \\ \hline

\begin{tabular}{@{}c@{}}\rev{WDC} \\\rev{Products}\end{tabular} & \rev{DITTO (128)}        & \rev{19.71 $\pm$ 0.00} & \rev{36.96 $\pm$ 0.00} & \rev{25.71 $\pm$ 0.00} & \rev{1.19 $\pm$ 0.31} & \rev{50.38 $\pm$ 4.02} & \rev{2.33 $\pm$ 0.59} & \rev{0.01 $\pm$ 0.00} & \rev{72.59 $\pm$ 2.21} & \rev{9.02 $\pm$ 1.67} & \rev{16.03 $\pm$ 2.69} & \rev{0.84 $\pm$ 0.00} & \rev{\textbf{31 sec}} \\
& \rev{DITTO (256)}        & \rev{19.71 $\pm$ 0.00} & \rev{36.96 $\pm$ 0.00} & \rev{25.71 $\pm$ 0.00} & \rev{20.34 $\pm$ 0.06} & \rev{39.97 $\pm$ 0.86} & \rev{26.96 $\pm$ 0.15} & \rev{0.01 $\pm$ 0.00} & \rev{74.14 $\pm$ 2.89} & \rev{18.06 $\pm$ 2.72} & \rev{28.96 $\pm$ 3.30} & \rev{0.85 $\pm$ 0.04} & \rev{32 sec} \\
& \rev{DistilBERT (128)-ALL}           & \rev{39.64} $\pm$ \rev{0.57 }        & \rev{65.27} $\pm$ \rev{1.15}       &  \rev{49.32} $\pm$ \rev{0.50}      & \rev{7.47} $\pm$ \rev{4.78}  & \rev{71.4} $\pm$ \rev{1.56} & \rev{13.03} $\pm$ \rev{7.75} & \rev{0.43} $\pm$ \rev{0.01} & \rev{35.54} $\pm$ \rev{1.29} & \rev{57.93} $\pm$ \rev{0.47} & \rev{\textbf{44.04 $\pm$ 1.06}}  & \rev{0.53} $\pm$ \rev{0.01} & \rev{40 sec }         \\ 
\hline 

\end{tabular}
}
\end{table*}

\subsection{Entity Group Matching with Blocking and GraLMatch}

\subsubsection{\textbf{Companies Datasets}}

Table \ref{tab:group_matching_table} shows the precision, recall and F1 scores of the entity group matching experiment of \emph{both real and synthetic companies datasets}.

Let us first analyze the \emph{pairwise matching performances} \rev{on pairs produced by blocking (first column of Table \ref{tab:group_matching_table})}. For all models, the recall and precision of the pairwise match predictions are slightly lower than those obtained during fine-tuning. The lower recall results from a portion of true pairs not being selected by blocking as candidate pairs. The lower precision is due to the fact that the negative pairs among the candidate matches are much more difficult to classify than the randomly sampled negative pairs used during fine-tuning. 

Next, we analyze the \emph{entity group matching performance} (second column of Table \ref{tab:group_matching_table}). As illustrated in Figure \ref{fig:Sets_of_Matches_considered}, false positive pairwise predictions greatly affect the entity group matching performance, since they connect different groups of entities leading to numerous false positive predictions. This phenomenon is reflected by the \emph{Pre Graph Cleanup} Precision and Cluster Purity scores, which evidence the necessity of removing false positive pairwise predictions to achieve a good entity group matching. Additionally, it is more pronounced on the synthetic companies because they are much more numerous (see Table \ref{tab:blockings_datasets}) and thus contain considerably more records sharing common terms ("hi-tech", "networks", "energy", "resources", geographical terms etc.) in their names\footnote{This is also the case in the complete real database, but the set of labeled records is considerably smaller in size and thus this phenomenon is not as significant.} which make false positive predictions more likely.

The \emph{Post Graph Cleanup} scores show how false positive predictions can be detected and removed with graph-based techniques such as Algorithm \ref{alg:Graph Cleanup}. The recall scores are lower than the \emph{Pre Graph Cleanup} ones because the GraLMatch Graph Cleanup also removes true positive matches, but the precision, F1 and Cluster Purity scores show that without the GraLMatch Graph Cleanup, a good entity group matching is not achieved.

Comparing the scores for the different models, we can see that DistilBERT(128)-ALL has a higher pairwise recall but lower precision than DistilBERT(128)-15K. This lower precision forces the GraLMatch Graph Cleanup to remove more predictions, which causes it to end up with a lower F1 score. DistilBERT(128)-15K has a lower recall due to being trained on fewer samples but achieves a higher precision. This higher precision means the GraLMatch Graph Cleanup does not need to remove as many predictions and makes it end up with the highest final F1 score.

Conversely, all DITTO variants consistently achieve a low pairwise precision score on both real and synthetic companies, leading to low \emph{Pre Graph Cleanup} F1 and Cluster Purity scores. On the real companies, the GraLMatch Graph Cleanup works surprisingly well for DITTO and it achieves the highest F1 scores but this is not the case for synthetic companies. Given that the synthetic companies dataset is closer in size to the complete real companies dataset (see Table \ref{tab:blockings_datasets}) and contains more challenging record groups to match (real companies mostly contain record groups obtained via identifiers which are easy to match), we come to the conclusion that having a high pairwise precision is the most determining factor in achieving a good entity group matching for large numbers of records.

DistilBERT(128)-15K is the model reaching the highest precision and the best \emph{Post Graph Cleanup} F1 score for synthetic companies. This exemplifies the fact that training with more data (and thus investing more in labelling) and/or carrying out fine-tuning optimizations do not necessarily guarantee a good entity group matching performance, where precision is the most important factor.

\rev{Regarding the sensitivity of Algorithm \ref{alg:Graph Cleanup}, we can observe how DistilBERT (128)-ALL-BC obtains a lower \emph{Post Graph Cleanup} Recall score than DistilBERT (128)-ALL, since the \textit{Minimum Edge Cut} removes more true positive edges than Algorithm \ref{alg:Graph Cleanup} albeit in a slightly shorter time. In turn, DistilBERT (128)-ALL-BC achieves the same scores as DistilBERT (128)-ALL with only a slightly longer running time because the \textit{Edge Betweenness Centrality} selects the same edges for removal as the \textit{Minimum Edge Cut.} DistilBERT (128)-ALL ($\frac{1}{2}\gamma$) obtains a half-way result between both previous setups, confirming our intuition that generally the \textit{Minimum Edge Cut} is the faster but less accurate edge removal technique. The specific threshold values we choose prove to be a good compromise of accuracy vs speed although the small time differences  signify that the running time of the experiment is dominated by the inference of the language models. Finally,  the similar final Post Graph Cleanup F1 scores indicate that Algorithm \ref{alg:Graph Cleanup} robustly achieves similar results with different choices of thresholds.}

Regarding the groups affected by \textit{data drift} events, record pairs presenting different textual attributes (names, descriptions) are likely being predicted as non-matches, since fine-tuning teaches the models to predict pairs based on text alignment. Consequently, scores largely represent how good models are at matching non-edge case groups. In a real setting, however, most edge case groups can be easily identified as they present matching identifiers but are predicted as non-matches by the language models. Their treatment will require additional knowledge sources, to identify the type of \textit{data drift} event, and should be done on a case-by-case basis.

\subsubsection{\textbf{Securities Datasets}}

% Pairwise Matching with Blocking

We now analyze the entity group matching results on the security datasets. Similar to the matching of companies, the pairwise matching performance (see first column of Table \ref{tab:group_matching_table}) is generally worse than the performance on the fine-tuning test pairs (see Table \ref{tab:test_pairwise_table}). We observe on the real security dataset that both DITTO models predominantly predict any pair to be a match, visible in the low precision while attaining a high recall, whereas DITTO (256) performed well on its test set. The synthetic dataset features more randomized identifiers and many pairs which cannot just be matched by finding an identifier overlap. Because of this, all models achieve precision above 95\%, while their recall is around 53\%, with the exception of DistilBERT(128)-15K at 57\%.

% Graph Cleanup (Transitive Matches Added)

During the pre graph cleanup (see second column of Table \ref{tab:group_matching_table}) we see a slight increase in recall at the cost of precision, though the securities are forming smaller clusters and thus do no exhibit as many transitively matched false positives. Because of this, the reduction in precision is much less severe, compared to the company datasets. Furthermore, we observe the same pattern as on the synthetic company dataset, where the DistilBERT(128)-15K variant trained on less data outperforms the DistilBERT(128)-ALL model.

% Graph Cleanup (After all is applied)
After cleaning up the prediction graphs on the real security dataset, both DITTO models reach lower metrics compared to the pairwise matching performance with F1-Scores around 20\%. This clearly shows how the initial predictions have failed to learn any useful pattern. The DistilBERT(128)-ALL model by comparison reaches a F1-Score of 98.86\%, further indicating that the encoding structure of DITTO is not ideal for identifier-centric data. On the synthetic security dataset, the methods achieve a comparable precision, but the DITTO model needs to be double in size to match the DistilBERT(128)-15K model's performance in recall. DistilBERT(128)-ALL here reaches the highest F1-Score of 72.11\%, though its precision is lower than the other models.

%\todo[inline]{In the securities datasets, where we already start off with high cluster purity (because we only use the id overlap blocking), the graph clean up doesn't change much the entity group matching. This makes sense, the id overlap blocking already leads to very small connected components and thus the graph clean up removes much fewer edges. In some cases (synthetic securities/Ditto (128)) it even makes the final result worse. Additionally, the final recall is very low because we block all pairs without id overlaps since we would need issuer information to match those, this class of matches are also ignored in the real securities.}

\subsubsection{\textbf{WDC Products}}

\rev{Finally, we analyze the entity group matching results on the WDC Products dataset, presented in the last rows of Table \ref{tab:group_matching_table}. As in the other datasets, the pairwise scores are worse than those obtained during fine-tuning due to the same reasons i.e. some true pairs being discarded by the blocking and the blocking candidate pairs being more difficult to classify than the test pairs evaluated during fine-tuning. In terms of entity group matching, we can again observe how the \emph{Pre Graph Cleanup} Precision scores drop due to the effect of false positive pairwise predictions. The model with the highest pairwise precision, DistilBERT(128)-ALL, achieves the highest \emph{Post Graph Cleanup} F1 score whereas DITTO (128) and DITTO (256), which present much lower pairwise precisions, both achieve lower F1 scores. Contrary to other datasets however, DITTO (256) achieves the best \emph{Pre Graph Cleanup} F1 score but not the best \emph{Post Graph Cleanup} F1 score. This is likely due to our Graph Cleanup method, which is not fit to matching settings where record groups of heterogeneous sizes have to be discovered such as the one this dataset presents. Using a better suited Graph Cleanup method should revert this phenomenon.}

\section{Conclusions}
In this paper, we have discussed the particularities of the multi-source Entity Matching problem, that we refer to as \emph{entity group matching}, where the challenge is to assign records, belonging to the same real-world entity, to the same group.  We have presented two new benchmark datasets made up of companies and financial securities records, inspired by a real-world use-case. We have \rev{introduced the concept of \emph{transitively matched records} and} illustrated how false positive pairwise predictions affect the group assignment of large numbers of records \rev{leading to considerable numbers of false transitive matches.} We have shown how these false \rev{pairwise} predictions can largely be corrected via the use of graph-based algorithms such as the one we propose, \emph{GraLMatch}. 

Moreover, our experiments have shown that fine-tuning DistilBERT, a Transformer-based model with relatively few parameters, on a limited number of records yields a better out-of-sample performance than fine-tuning with more samples and/or with different fine-tuning optimizations. As our experiments illustrate, precision is the key to achieving a good entity group matching especially with large volumes of records. 

\rev{Our approach \emph{addresses and solves a pressing real-world problem in the financial industry}. Companies operating in this sector spend significant time on obtaining data from different data vendors to ensure comprehensive coverage across different instruments and geographies. The costs associated with these data licenses can reach millions of dollars for comprehensive data feeds from major providers like Bloomberg, Thomson Reuters (Refinitiv), etc. Our proposed approach will allow companies to have one-stop-shop access to financial data, which underpins almost every aspect of their operations.}

\label{Conclusions/Further Work}

\begin{acks}
We thank Damian Tschirky and Gabriele Von Planta, both formerly of Move Digital AG, for their valuable insight during the course of the project. The work was funded by Innosuisse as an innovation project under the project number \texttt{54383.1 IP-ICT}.
\end{acks}

% %To break the references on their own column
% \vfill\null
% \pagebreak

\bibliographystyle{ACM-Reference-Format}
\bibliography{main}

\end{document}